\documentclass[%
 reprint,
 amsmath,amssymb,
 aps,usenames,dvipsnames
]{revtex4-1}

\usepackage{tikz}
\usepackage{rotating}
\usepackage{graphicx}
\usepackage{dcolumn}
\usepackage{bm}
\usepackage{gensymb}
\usepackage{hyperref}
\usepackage{longtable}
\usepackage{booktabs,array}
\usepackage{hhline}
\usepackage[usename]{pstcol}

\begin{document}

\preprint{APS/123-QED}

\title{On the Microscopic Level Density Models for Nuclei Near Z=28 Shell Closure}

\author{Surayya H.E$^1$}
\author{Jesmi Sunny$^1$}
\author{M.M Musthafa$^1$}\email{mmm@uoc.ac.in}
\author{C.V Midhun$^1$}\email{midhun.chemana@gmail.com}
\author{S.V Suryanarayana$^2$}
\author{Jyoti Pandey$^3$}
\author{A Pal$^2$}
\author{P.C Rout$^2$}
\author{S Santra$^2$}
\author{Antony Joseph$^1$}
\author{S. Ganesan$^4$} 
\affiliation{$^1$ Department of Physics,University of Calicut, Calicut University P.O Kerala, 673635 India}
\affiliation{$^2$ Nuclear Physics Division, Bhabha Atomic Research Centre, Mumbai 400085, India}
\affiliation{$^3$Inter University Accelerator Centre, New Delhi, Delhi 110067, India}
\affiliation{$^4$ Formarly Raja Ramanna Fellow, Bhabha Atomic Research Centre, Mumbai 400085, India}

\date{\today}

\begin{abstract}
\section*{Abstract}
A comprehensive test of level density models for explaining the decay of excited compound nuclei, $^{54}$Mn, $^{56}$Fe, $^{58}$Co, $^{60}$Ni, $^{61}$Ni and $^{63}$Cu, in the energy range of 28 - 36 MeV has been performed. The compound nuclei of interest in the desired ranges are populated using $^{6}$Li based transfer reactions. The proton decay spectrum for each excitation energy bins has been measured. The measured proton spectrum has been reproduced using statistical model calculations with different level density models. A variance minimised approach has been employed for analysing the prediction capability of different level density models. This approach has been converged to Gogny Hartree-Fock-Bogoliubov(HFB) microscopic level density model and which is attributed as the most accurate model for the desired nuclei.

\end{abstract}

\maketitle


\section{\label{sec:level1}introduction}
The nuclear reaction rates assessed from the neutron and proton spectra are required for nuclear astrophysics, heavy ion reactions and reactor applications. As most of the nuclei are non-existing, all the relevant data cannot be directly measured or easily predicted through theory. The level densities are important inputs in the statistical model of nuclear reactions, i.e, the fusion evaporation model which provide an easy solution \cite{ramirez2013nuclear}. It is well known that the level density parameter is energy dependent and that the shell structure has a significant impact on the level density at low excitation energies \citep{hilaire2004energy}. When excitation energies are large, the shell effect disappears \citep{aggarwal2023impact}.  A current issue in nuclear physics is the study of the fundamental properties of atomic nuclei at high excitation energies, large angular momenta, substantially deformed states and for nuclei which are off of the stability line. \citep{behkami2001nuclear}. In nuclear astrophysics, the r-process is responsible for the creation of approximately half of the atomic nuclei heavier than iron \citep{mumpower2016impact}. As we move towards the neutron dripline, it supresses r-process. The (n, $\gamma$) reaction is the most important one competing with the (n, 2n) reaction perhaps explained using level density information \citep{scielzo2011determining}.

In fusion reactors, the neutron induced reaction cross-section data are used for the evaluation of radiation damage to its components, mainly the production of H and He gases \citep{gilbert2012integrated} \citep{pandey2022estimation}. The reactor structural material, viz., stainless steel, have Cr, Fe, Ni, Mn and Co as its main constituents. The long-lived radio isotopes of these elements produced in the reactor participates in highly exo-thermic reactions which in turn lead to significant radiation damage issues and nuclear waste \citep{wallner2011production}. Therefore, it is important to study their cross-sections as well as proton emission channels. Phenomenological and empirical models work well for the stable or well-known reachable nuclei. Moving from the stable nuclides to weakly bounded region, the extrapolation and therefore the predictive capability of empirical or statistical models need to be thoroughly investigated. However, microscopic models are found to be considerably better in predicting over all regions including stable to weakly bounded systems \citep{sun2023microscopic}. The microscopic model includes a consistent treatment of the shell effects, pairing correlations, deformation effects and collective excitations. The microscopic approaches are obviously to be preferred in the collective enhancement on nuclear level density(NLD) and quasi bound nucleus in direct process \citep{zhao2020microscopic}.

NLD depends upon the nuclear structure and this variation is more evident in proton emission spectra, exhibited in the form of particle energies. Conventional level density model fails to explain the structural dependences. Hence the total emission spectra are best suited to figure out the reliability of microscopic level density models in analysing the energy distribution and multiplicity of outgoing particles. The evaporation method is particularly suitable at higher excitation energies and less sensitive to spin and parity distributions \citep{roy2011study}. It is to be noted that the prediction of different level density model differ for same nuclear system. Hence the acceptability testing and optimization of the microscopic level densities are important to explain the nuclear reactions. The best level density model will be capable of predicting the cross sections for large number of systems in wide range of mass and energy. In the present work, we have attempted to optimize the nuclear level density models that can predict the measured proton spectra of nuclides near Z=28. Accordingly, the proton spectra from the compound nuclei $^{54}$Mn, $^{56}$Fe, $^{58}$Co, $^{60}$Ni, $^{61}$Ni and $^{63}$Cu over the excitation energy range of 28-36 MeV are  measured and analysed using various level density models. The details of the measurement and analysis are described in the following sections.

\section{Theoretical level density models}
There are various NLD models proposed for treating the nuclear reaction mechanisms. Most of the level density models are empirical or phenomenological. Some of the most used models are discussed below.

\subsection{Fermi Gas Model:}

Fermi Gas Model \citep{mengoni1994fermi} is based on the assumption that the nucleons, which are not interacting each other, are confined in a three dimensional barrier having size of the nucleus. This confinement of nucleons produces occupancies, according to the possible eigen value, based on the Pauli’s exclusion principle. This assumes  that all the energy levels are single particle states. Based on this picture, the level density is expressed as \citep{mengoni1994fermi}.
\begin{equation}
\rho(U)=\frac{\sqrt{\pi}e^{2\sqrt{aU})}}{12{\sigma}a^{1/4}{U}^{5/4}}
\end{equation}
where \textbf{\emph{a}} is the level density parameter, \textbf{U} is the excitation energy of the residual nucleus, and \textbf{$\sigma$} is the spin cut off parameter.

The Fermi Gas Model(FGM) is based on an over simplified assumption that nucleons are non-interacting with
each other and are bound by the nuclear meanfield,
neglecting residual interactions and many body effects.
Consequently, this model cannot account for the effects
of shell closure, pairing correlations, coherent excitations
of collective nature and discontinuities in the structure
on level densities. Hence, it fails to explain the level
density distribution in cases where the above interactions are prominent. \citep{schmidt2012inconsistencies}.
\subsection{Constant Temperature Model:}
Constant Temperature Model(CTM) \citep{zelevinsky2018constant} treats the excited nucleus as a superheated liquid and particle evaporations as the phase transitions. Hence the level densities are treated as the density of states in the superheated liquid, which is independant of the present temperature. The density of states are considered based on the quantum distribution function of fermions. Based on this picture, the density of states are expressed as
\begin{equation}
\rho(E) = \frac{1}{T}e^{\frac{E-{E_0}}{T}}         
\end{equation}
where \textbf{T} is the effective nuclear temperature and \textbf{E${_0}$} is the adjustable reference energy level that can be varied to match the experimental discrete levels, while \textbf{E} is the actual nuclear excitation energy
counted from the ground state.
The exponential growth, eqn(2), cannot continue for too high energies as this would contradict to the condition of thermodynamic equilibrium.
 
\subsection{Back-Shifted Fermi Gas Model:}
It is the most popular approach to estimate the spin-dependent NLD. Mostly the Back-shifted Fermi-gas(BFGM) \citep{dilg1973level} description, assuming an even distribution of odd and even parities, is used. In the Back-shifted Fermi Gas Model, the pairing energy is treated as an adjustable parameter and the Fermi gas expression is used all the way down to 0 MeV. Hence for the total level density we have
\begin{equation}
\rho(U,J,\pi) = \frac{1}{2}\frac{2J+1}{2\sqrt{2\pi}\sigma^3}exp[\frac{-(J+1/2)^2}{2\sigma^2}]\frac{\sqrt{\pi}}{12}\frac{exp[{2\sqrt{aU}}]}{a^{1/4}U^{5/4}}
\end{equation}

where $\sigma$ is the spin cut-off parameter, \textbf{\emph{a}} is the level density parameter and \textbf{U} is the effective excitation energy.

The state in the Back-shifted Fermi Gas Model (BFGM) is that at high energies, the excitation energy of the Fermi gas expression includes a shifting parameter. Large uncertainities are expected in the BFGM prediction of NLD, especially when extrapolating to very low (a few MeV) or high energies $(U\geq{15MeV})$ and/or to nuclei far from the valley of $\beta$-stability.
\subsection{Generalized Superfluid Model:}
 The Generalized Superfluid Model(GSM) \citep{ignatyuk1993density} takes superconductive pairing correlations into account, according to the Bardeen-Cooper-Schrieffer(BCS) theory. The phenomenological version of the model is characterized by a phase transition from a superfluid behaviour at low energy, where pairing correlations strongly influence the level density to a high energy region, which is described by the FGM. At low energies it resembles CTM. To take pairing correlations and collective effects into account, it is necessary to use the following
expression for the total level density:
\begin{equation}
\rho(U) = \rho_{quasiparticle}(U)K_{vibr}(U)K_{rot}(U),
\end{equation}
where, $k_{vib}$ and $k_{rot}$ are called the rotational and vibrational enhancement factors, respectively.
\begin{equation}
\rho^{tot}_{GSM}=\frac{1}{\sqrt{2\pi}{\sigma}}\frac{\sqrt{\pi}e^{2\sqrt{aU}}}{12a^{1/4}U^{5/4}}
\end{equation}
where,\textbf{U}, \textbf{\emph{a}} and \textbf{$\sigma$} are the effective excitation energy, level density parameter and spin cut-off parameter respectively. 

Level density parameters for the GSFM, which takes into account collective enhancement of the nuclear level density, in addition to shell and superfluid effects.

\subsection{Microscopic Level Density Models:}
The microscopic level density model includes Skyrme HF-Bogoliubov(HFB)\citep{minato2011nuclear}, Gogny HFB \citep{goriely2009first}, Temperature-dependent Gogny HFB model \citep{hilaire2012temperature}. Skyrme HFB plus BCS method treats shell effects and pairing correlation self-consistently. Gogny interaction has ability to predict low-energy quadrupole collective levels, both rotational and vibrational. It enables microscopic description of energy dependent shell and parity. The Temperature dependent Gogny HFB has been implemented for accounting the  transition from deformed to spherical shape at increasing excitation energies.
 
The level density due to shifted Bethe formula \citep{behkami2002level} based on statistical approach is given by,
\begin{equation}
\rho(E)=\frac{\exp[2\sqrt{a(U-S)}]}{\sqrt{48}a^{1/4}(U-\delta)}
\end{equation}

Where \textbf{S} is the entropy of the compound nuclear system, \textbf{a} is the level density parameter, $\delta$ is the  energy shift pararmeter and \textbf{U} is the excitation energy.

The Skyrme interaction is introduced for accounting the structural description of the nucleus, at the particular excitation energies. This accounts for the density- dependent nucleon-nucleon interaction and thereby the energy levels with the effect of collective models are generated. This is mathematically deduced as a solution of Hartree-Fock(HF) equations \citep{goodman1981finite}, \citep{egido1986temperature}. This quantum mechanical approach provides energy, spin, and parity-dependent NLDs, proposed by Hilaire and Goriely \citep{goriely2008improved}, at low energies which describes the nonstatistical limit. The nonstatistical approach in the case of stationary states of even-even nuclei is given by \citep{chen2022three},

\begin{equation}
\rho_q (r) = \sum_{\alpha \in q} \sum_{s} v_{\alpha}^2 |\psi_\alpha (r,s)|^2 ,
\end{equation}
\begin{equation}
\tau_q (r) = \sum_{\alpha \in q} \sum_{s} v_{\alpha}^2 |\Delta\psi_\alpha (r,s)|^2 ,
\end{equation}
\begin{equation}
J_q (r) = -i \sum_{\alpha \in q} \sum_{ss'} v_{\alpha}^2 \psi_\alpha ^2(r,s) \Delta \times \sigma_ss' \psi_\alpha (r,s')
\end{equation}

where $\psi_\alpha$ is the spin-parity wave functions with fractional occupation amplitudes, $v_{\alpha}$. q stands for the index of protons or neutrons. s and $s^{\prime}$ are the labels of two spinor components of wave functions which takes the value ±1/2. 
Here, the energy dependence is limited to the local particle density $\rho_q$, the kinetic energy density $\tau_q$, and the spin-orbit density $J_q$.
By definition, the non statistical limit cannot be described by the traditional, statistical formulas. This nonstatistical behavior can have a significant impact on the cross sections, particularly sensitive to spin-parity distributions. The solution of HF equations is a computationally challenging process to generate the level density information. Hence, the level densities for most nuclei at particular excitation energies are made available in tabular format, based on parametrization of Goriely and Gogny and are included in RIPL \citep{capote2009ripl}.

\section{Methodology}
The compound nuclei $^{54}$Mn, $^{56}$Fe, $^{58}$Co, $^{60}$Ni, $^{61}$Ni and $^{63}$Cu  were populated at overlapping excitation energies  rangeing over 28 - 36 MeV through $^{6}$Li induced transfer reactions. The evaporated proton spectra from the excited nuclei has been measured. The measured spectra is converted to decay probabilities, by normalizing with the number of compound nuclei produced. The prediction capability of level density models has been validated by analysing the variation of calculated and experimental spectrum. The proton spectra from the same compound nuclei populated are theoretically calculated using conventional and microscopic level density formalisms. 
       
\section{Experimental details}
The experiment was carried out at BARC-TIFR Pelletron-Linac facility in Mumbai using $^{6}$Li beam . The self-supporting target of $^{59}$Co (abundance $\approx$ 100$\%$), $^{nat}$Fe (abundance of $^{56}$Fe $\approx$ 92$\%$) of thickness 700µg/cm$^{2}$ and $^{nat}$Cr (abundance $^{52}$Cr $\approx$ 84$\%$) of thickness $\approx$ 500 $\mu$g/cm$^{2}$ were prepared using the thermal evaporation technique. The Co and Cr targets were bombarded with $^{6}Li$ beam at incident energy E$_{lab}$ = 40.5 MeV. While, Fe target was bombarded at incident energy E$_{lab}$ = 35.9 MeV.
The transfer reactions populated corresponding to the compound nuclei and their excitation energies are presented in Table.\label{Tab:2}\ref{Tab:2}.

\begin{table}[h]
\caption{Transfer reactions investigated in the present experiment, the compound nucleus formed and their corresponding excitation energy ranges.}
\begin{tabular}{c c c}

\hline\hline
Transfer Reaction & $\hspace{1mm}$  Compound Nucleus & $\hspace{1mm}$ Excitation Energy\\
                  &                                  &                 Ranges\\
\hline

$^{59}$Co($^{6}$Li,$\alpha$) &  $^{61}$Ni  &  29.55 - 32.55 \\

$^{59}$Co($^{6}$Li,d)  &   $^{63}$Cu  &  30.29 - 32.91\\

$^{56}$Fe($^{6}$Li,$\alpha$) &  $^{58}$Co & 28.90 - 31.90\\

$^{56}$Fe($^{6}$Li,d) &   $^{60}$Ni &  31.64 - 34.43\\

$^{52}$Cr($^{6}$Li,$\alpha$) &  $^{54}$Mn  &  29.80 - 32.80\\

$^{52}$Cr($^{6}$Li,d)   &   $^{56}$Fe  & 33.09 - 35.85 \\
\hline\hline
\end{tabular}
\label{Tab:2}
\end{table}

\begin{figure}[hbtp]
\includegraphics[scale=0.45,trim={7cm, 2cm, 0cm, 2cm},clip]{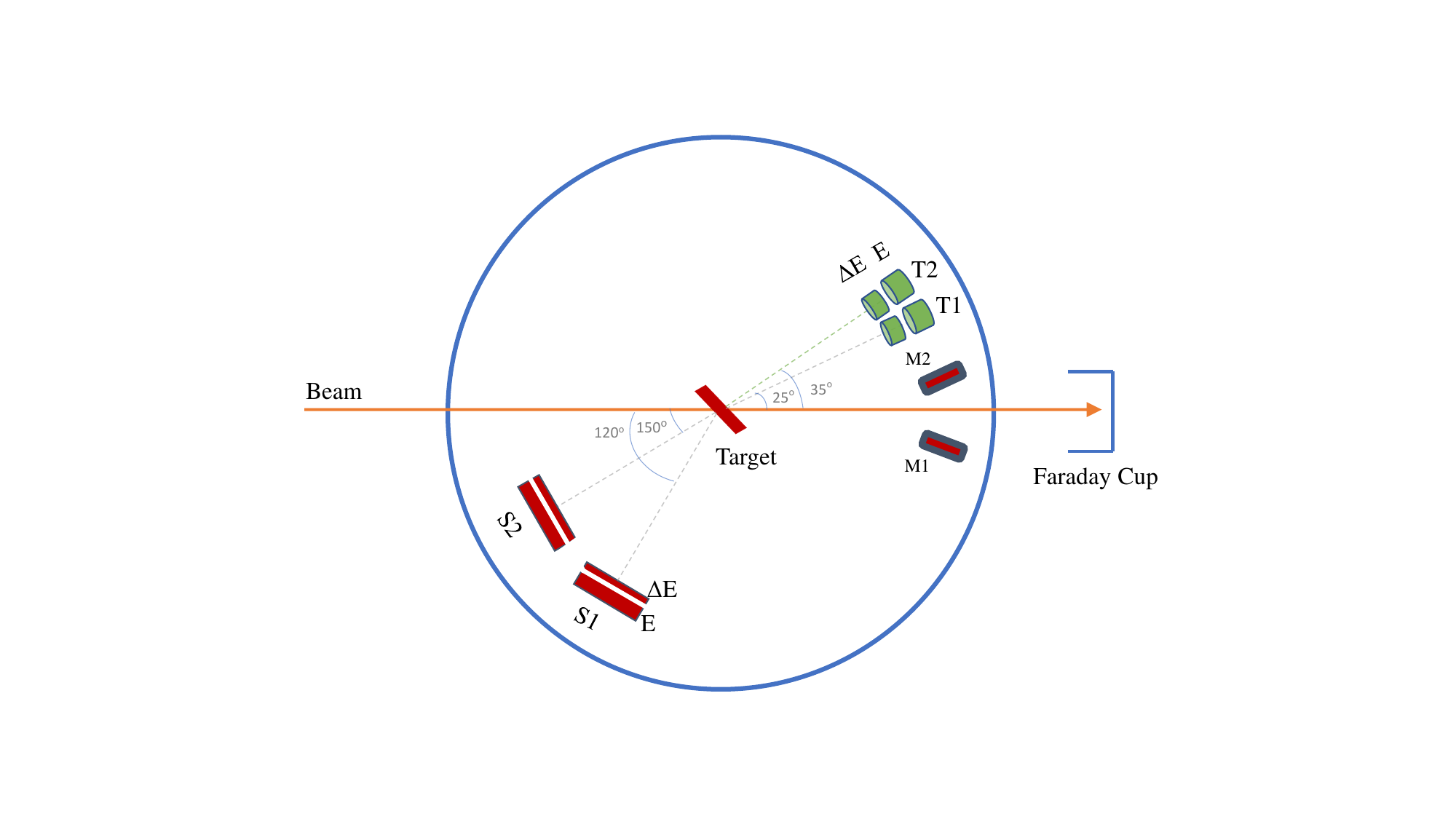}
\caption{\label{fig:f1}The schematic diagram of experimental setup}
\end{figure}

\begin{figure}[hbtp]
\includegraphics[width=\columnwidth]{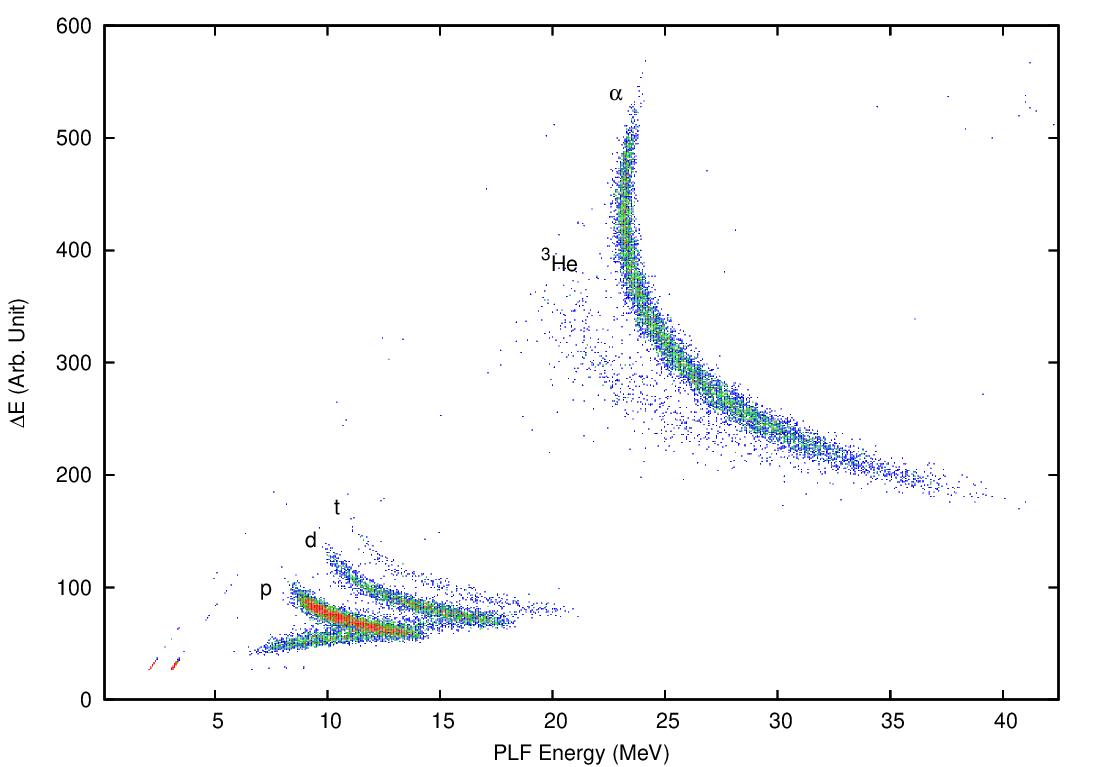} 
\caption{\label{fig:2}Typical E-$\Delta$E spectrum obtained from telescope T2 for $^{6}$Li+$^{59}$Co reaction at E$_{lab}$ = 40.5 MeV}
\end{figure}

\begin{figure}[hbtp]
\includegraphics[width=\columnwidth]{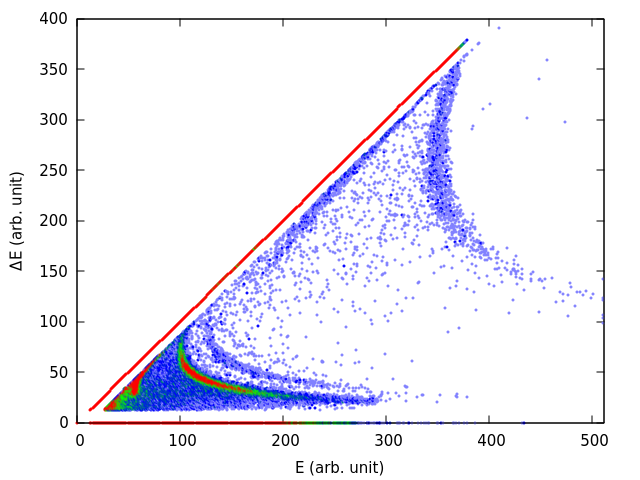}
\caption{\label{fig:f3}Typical E-$\Delta$E  spectrum obtained from one of the 32 $\Delta$E-E strip combinations for  $^{6}$Li + $^{56}$Fe reaction}
\end{figure}

\begin{figure}[hbtp]
\includegraphics[width=\columnwidth]{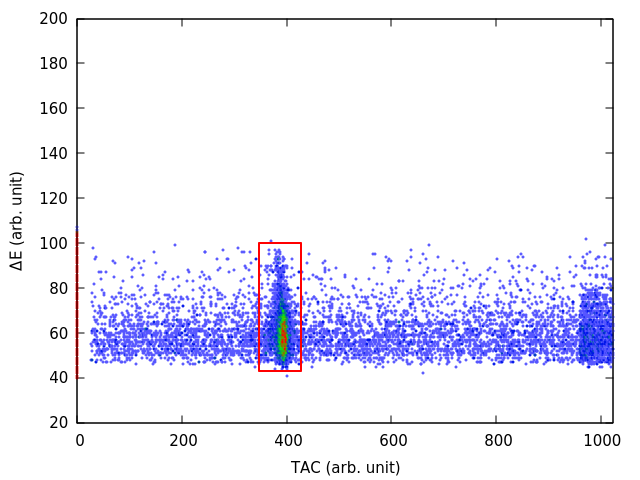}
\caption{\label{fig:f4}A typical two-dimensional plot of TAC versus PLF deuteron energy for $^{6}$Li+$^{56}$Fe reaction}
\end{figure}

To identify the projectile-like fragments (PLFs), two silicon detector telescopes (T1 and T2), having 150$\mu$m and 100$\mu$m $\Delta$E-E pairs, were mounted at 25$^0$ and 35$^0$ from the beam direction around the gracing angle of the transfer reaction. Two large area strip detector telescopes (S1 and S2) are placed in the backward angles to detect the evaporated particle from the compound nucleus covering an angular range of 110$^0$-130$^0$ and 140$^0$-160$^0$. These strip telescopes consist of two Si strip detectors having an active area $\approx$50 mm×50 mm placed back-to-back ($\Delta$E-E) with thickness of $\Delta$E $\approx$ 60 $\mu$m and E $\approx$ 1500 $\mu$m. Each detector has 16 vertical strips of size 3.1 mm×50.0 mm.The schematic diagram of experimental setup is shown in Fig.\label{f1} \ref{fig:f1}. In order to isolate the random coincidences in strips, due to the higher active area, the time correlations between PLFs and the evaporated particles are recorded through a Time to Amplitude Converter (TAC). A typical E-$\Delta$E spectrum obtained from telescope T2 for $^{6}$Li+$^{59}$Co reaction at beam energy of 40.5 MeV is illustrated in Fig.\label{f2} \ref{fig:2}, where all the PLFs like p, d, t  $^{3}$He and $^{4}$He are uniquely identified.
A typical two-dimensional $\Delta$E versus E spectrum obtained from one of the 32 $\Delta$E-E strip combinations for  $^{6}$Li+$^{56}$Fe reaction is shown in Fig.\label{f3} \ref{fig:f3}. A typical two-dimensional plot of deutron gated TAC versus PLF $\Delta$E, for $^{6}$Li+$^{56}$Fe reaction is shown in Fig. \label{f4} \ref{fig:f4}. The telescopes were calibrated using the known energies of the excited states of $^{16}$O* and $^{14}$N* formed in the experiment $^{12}$C($^{6}$Li,d)$^{16}$O* and $^{12}$C($^{6}$Li,$\alpha$)$^{14}$N*. Pu-Am $\alpha$ source was also used to calibrate the strips.

\section{Data Analysis}
The compound nuclei are populated at overlapping excitation energies in the range of 28 - 36 MeV corresponding to the PLF alpha energies, E$_{\alpha}$= 23.5 to 18.5 MeV and PLF deuteron energies, E$_{d}$= 12.5 to 8.5 MeV. The excitation energies of each event has been estimated by 

\newpage
\onecolumngrid
\begingroup

\begin{figure}[h]

\centering
\vspace*{-0.7cm}
\hspace*{-0.9cm}
\includegraphics[width=20cm,height=24cm]{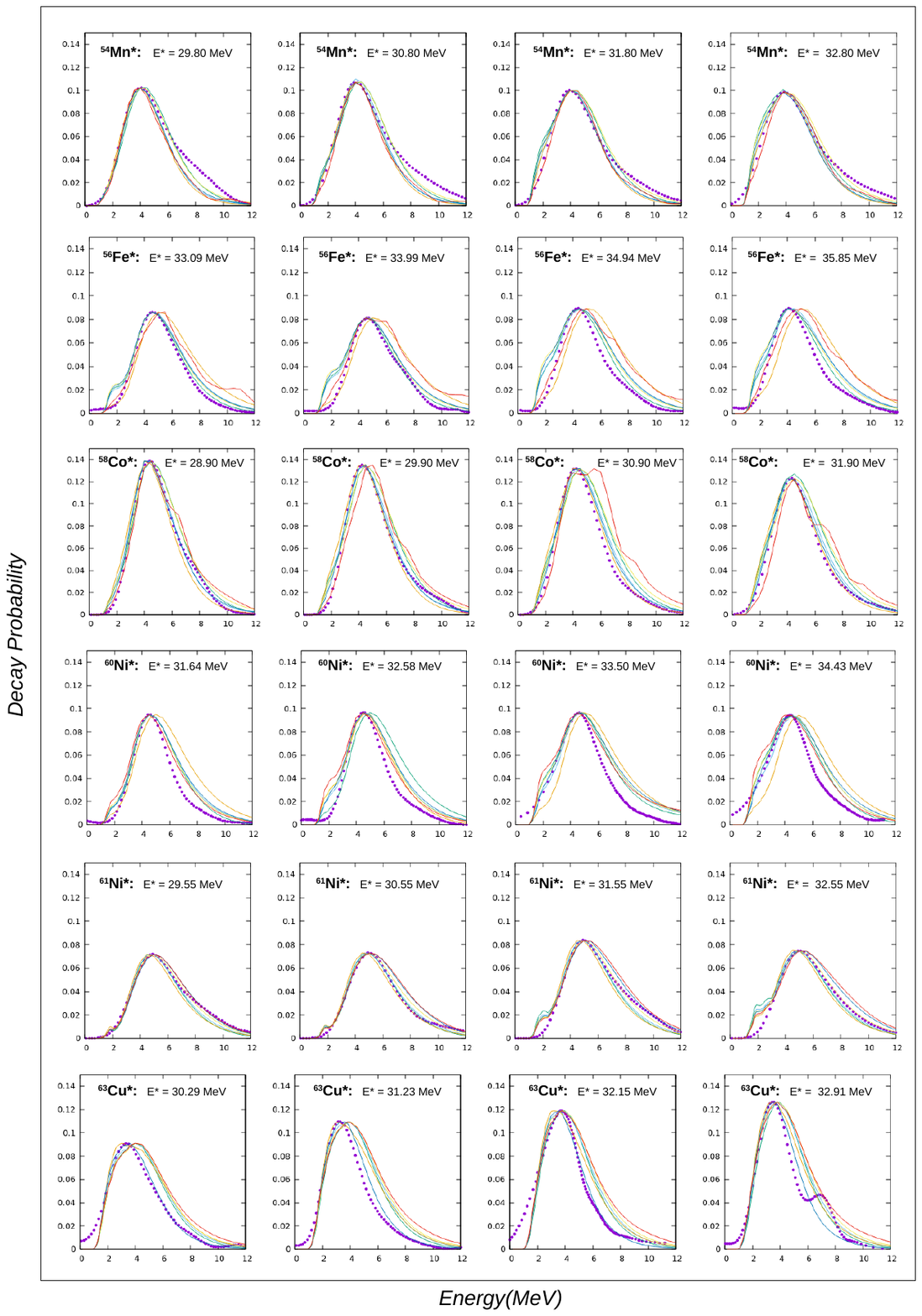} 
\caption{The measured proton decay spectra of $^{54}$Mn, $^{56}$Fe, $^{58}$Co, $^{60}$Ni, $^{61}$Ni and $^{63}$Cu compound nuclei for different excitation energies compared with theoretical level density models. In all the graphs, the dots represents the experimental data. Among the continuous lines; green, blue, orange, yellow, dark blue and red colors indicates CTM+FGM, BFGM, GSM, Skyrme HFB Model, Gogny HFB Model and Temperature-dependant Gogny HFB Model respectively.}

\label{fig:f5}
\end{figure}
\endgroup
\clearpage

\newpage
\onecolumngrid
\begingroup

\begin{figure}[h]
\centering
\includegraphics[width=18cm,height=9.5cm]{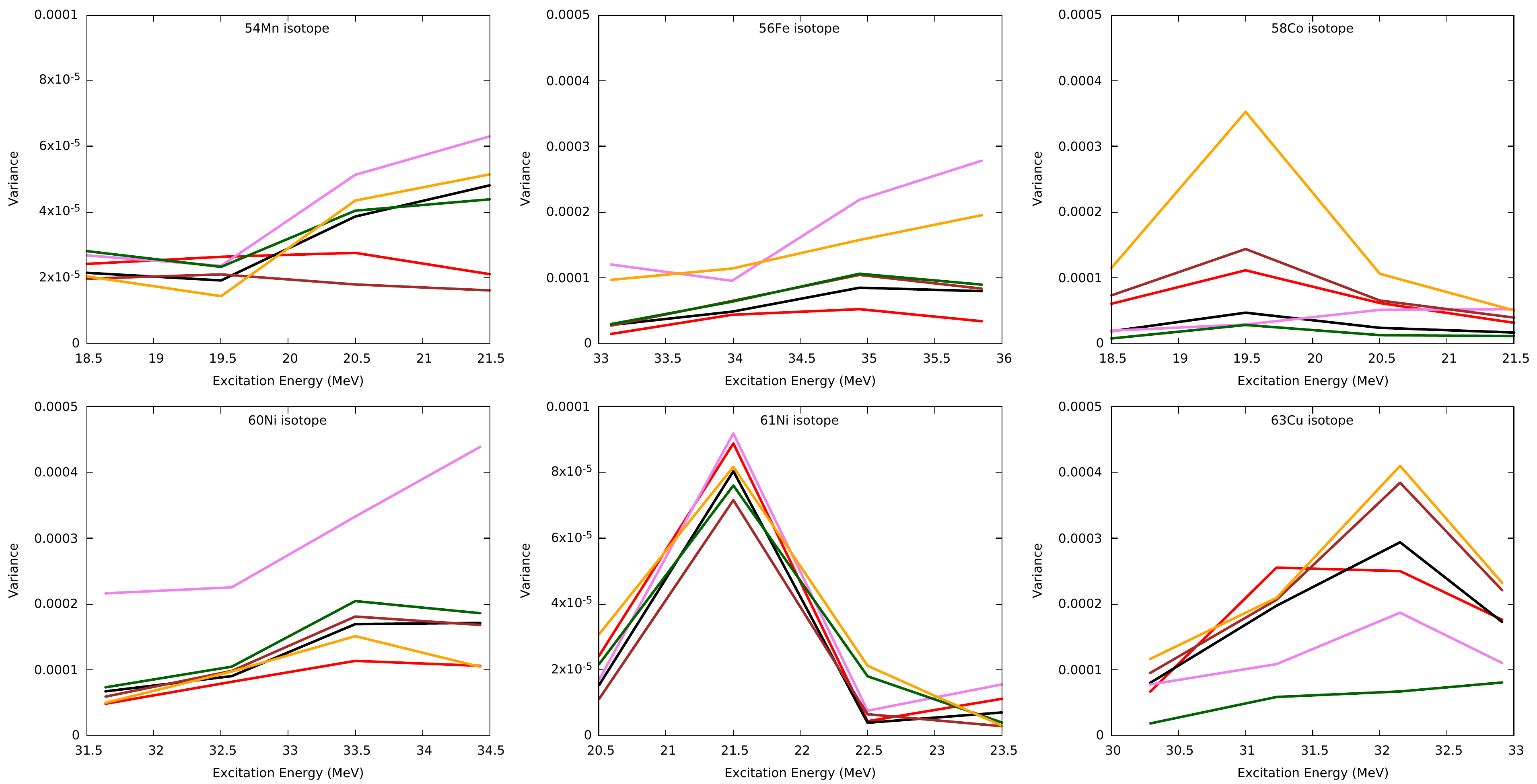}

\caption{The variance of each theoretical level density model from the experimental data for different excitation energies of each isotopes $^{54}$Mn, $^{56}$Fe, $^{58}$Co, $^{60}$Ni, $^{61}$Ni and $^{63}$Cu. The curves with colors red, black, violet, brown, green and orange represents CTM+FGM, BFGM, GSM, Skyrme HFB Model, Gogny HFB Model and Temperature-Dependent Gogny HFB Model respectively.}

\label{fig:f6}
\end{figure}
\endgroup
\twocolumngrid
\clearpage

two body kinematics. The evaporated proton events, from the excited compound nucleus has been obtained by generating coincidence between PLF and proton events in the strip detectors. Further the TAC gate  isolates the random events. The deuteron/alpha gate and TAC gate on the strip detector give the evaporated proton spectra from the compound nuclei. The proton decay probabilities of compound nuclei produced in the transfer reactions are obtained using the relation;
\begin{equation}
\Gamma^{CN}_{p}(E_{x}) = \frac{N_{i-p}(E_{x})}{N_{i}(E_{x})}
\end{equation}                                                 
The subscript ‘i’ in Eq. (1) denotes the transfer PLF corresponding to the populated compound nuclei of interest. N$_{i}$ and N$_{i-p}$ denote the singles and coincidence counts, at excitation energy E$_{x}$, respectively. The proton decay spectrum has been generated for 1 MeV steps of excitation energies. As the excitation energy is above the multiparticle emission threshold of the compound nuclei of interest, the proton spectrum is considered as orginated from (n,xp), where x is the accompanying particle such as n,p,$\alpha$ etc.

\section{Model Calculations and Comparison with Experimental Data}
The TALYS-1.96 code has been used to compare the experimental proton decay spectra evaporated from the compound nuclei $^{54}$Mn, $^{56}$Fe, $^{58}$Co, $^{60}$Ni, $^{61}$Ni and $^{63}$Cu. The total proton spectrum, considering all production channels such as p,$\alpha$p,2np etc. from the excited nuclei are calculated. The standard Hauser-Feshbach formalism \citep{hauser1952inelastic} was applied to calculate the compound nuclear decay.  The projectile-target combination corresponding to the transfer channel has been considered, and the potential parameters for the same has been attributed, to obtain the same $J^{\pi}$ states populated by the real experiment. The proton decay spectrum corresponding to the population (E$^{*}$,$J^{\pi}$) has been obtained, corresponding to Fermi Gas Model, Constant Temperature Model, Back-shifted Fermi Gas Model, Generalized Superfluid Model and Microscopic level density Models. The level density model parameters were obtained from RIPL-3 parameterization. Results using different level density models of TALYS-1.96 along with the present experimental data are compared. The variance criterion was used to attain optimum agreement between the experimental and calculated spectra.

\section{RESULTS AND DISCUSSION}
The experimentally obtained proton decay spectra from $^{54}$Mn, $^{56}$Fe, $^{58}$Co, $^{60}$Ni, $^{61}$Ni and $^{63}$Cu compound nuclei, for the excitation energies of 28-36 MeV are illustrated in Fig\label{fig} \ref{fig:f5}, along with the model predictions. In order to identify the suitability of each NLD model, variance is obtained using the equation:
\begin{equation}
variance= \frac{\sum{(x_{i}^{th}-x_{i}^{exp})^2}}{N}
\end{equation}
where, $\bf{x_{i}^{th}}$ and $\bf{x_{i}^{exp}}$ are the theoretical and experimental values of proton decay probability for a given proton energy; of a particular compound nuclei, respectively. And $\bf{N}$ is the total number of data.
 
The variance corresponding to the comparison of experimental and theoretical decay probability for each compound nuclei, with different excitation energies are illustrated in Fig\label{fig} \ref{fig:f6}. For each isotopes, different level density models produce a different agreement with the experimental data. Among these the Gogny HFB model based parameterization is more predictive for all the isotopes. Considering the phenomenological models, they are also reproducing the experimental data, based on the isotope specific parameterization from RIPL. However, the exact physics is lacking for these isotopes as the parameterizsation is more or less empirical. Considering, the nucleus as a statistical system, the exact physics of the nucleus, are getting heeled and the emperical parameterization is driving the emitted particle spectra. However, considering the microscopic models, they are accounting the real structural parameters based on the HFB calculations, which is more closer to the real physics situation inside the nuclei. Present study implies the acceptability of the HFB models, for explaining the particle spectrum, above multi-particle emission threshold in strongly bounded nuclei. For the multiparticle emissions, the shell gaps and level spin-parities are widely varying between the first particle emission and second particle emission. The present study sucessfully validates the compatibility of HFB based models for strongly and relatively weaker nuclei, having diffrent $J^\pi$ states.




\section{CONCLUSION}

The proton spectra from compound nuclei $^{54}$Mn, $^{56}$Fe, $^{58}$Co, $^{60}$Ni, $^{61}$Ni and $^{63}$Cu, populated by the transfer reactions, in the excitation energy range of 28-36 MeV, are measured. The data is utilised for validating the predictive power of various nuclear level density models, for nuclei around Z=28. Nuclear reaction code Talys-1.96, with RIPL-3 based parameterization, was used for the theoretical analysis. The present analysis shows that in the mass regions prominent to the shell effects and higher multiparticle emission probability, accounting the decay spectrum with higher accuracies require the microscopic level density calculations.

\section{ACKNOWLEDGEMENTS} 
Authors acknowledge the BARC-TIFR Pelletron-LINAC group for their support. This work is a part of M.Sc project of Surayya H.E. and Jesmi Sunny.
 

\def\bibsection{\section*{\refname}}
\bibliography{ref_ldm.bib}

\begin{thebibliography}{10}

\bibitem{ramirez2013nuclear}
A.~Ramirez, A.~Voinov, S.~Grimes, A.~Schiller, C.~Brune, T.~Massey, and
  A.~Salas-Bacci, ``Nuclear level densities of 64, 66 zn from neutron
  evaporation,'' {\em Physical Review C}, vol.~88, no.~6, p.~064324, 2013.

\bibitem{hilaire2004energy}
S.~Hilaire, ``Energy dependence of the level density parameter,'' {\em Physics
  Letters B}, vol.~583, no.~3-4, pp.~264--268, 2004.

\bibitem{aggarwal2023impact}
M.~Aggarwal, ``Impact of the quenching of shell effects with excitation energy
  on nuclear level density,'' {\em Nuclear Physics A}, vol.~1032, p.~122619,
  2023.

\bibitem{behkami2001nuclear}
A.~Behkami and Z.~Kargar, ``Nuclear level density at high spin and excitation
  energy,'' {\em Communications in Theoretical Physics}, vol.~36, no.~3,
  p.~305, 2001.

\bibitem{mumpower2016impact}
M.~R. Mumpower, R.~Surman, G.~McLaughlin, and A.~Aprahamian, ``The impact of
  individual nuclear properties on r-process nucleosynthesis,'' {\em Progress
  in Particle and Nuclear Physics}, vol.~86, pp.~86--126, 2016.

\bibitem{scielzo2011determining}
N.~Scielzo, J.~Burke, J.~Escher, S.~Collaboration, {\em et~al.}, ``Determining
  (n, $\gamma$) and (n, 2n) cross sections for radioactive isotopes using
  surrogate reactions,'' in {\em APS April Meeting Abstracts}, vol.~2011,
  pp.~X10--001, 2011.

\bibitem{gilbert2012integrated}
M.~Gilbert, S.~Dudarev, S.~Zheng, L.~Packer, and J.-C. Sublet, ``An integrated
  model for materials in a fusion power plant: transmutation, gas production,
  and helium embrittlement under neutron irradiation,'' {\em Nuclear Fusion},
  vol.~52, no.~8, p.~083019, 2012.

\bibitem{pandey2022estimation}
J.~Pandey, B.~Pandey, P.~Subhash, P.~Kanth, M.~Rajput, S.~Vala, R.~Makwana,
  S.~Suryanarayana, and H.~Agrawal, ``Estimation of production cross-sections,
  transmutation and gas generation from radionuclides (a~ 50--60) in fusion
  environment,'' {\em Applied Radiation and Isotopes}, vol.~184, p.~110163,
  2022.

\bibitem{wallner2011production}
A.~Wallner, K.~Buczak, C.~Lederer, H.~Vonach, T.~Faestermann, G.~Korschinek,
  M.~Poutivtsev, G.~Rugel, A.~Klix, K.~Seidel, {\em et~al.}, ``Production of
  long-lived radionuclides 10be, 14c, 53mn, 55fe, 59ni and 202gpb in a fusion
  environment,'' {\em J. Kor. Phys. Soc}, vol.~59, p.~1378, 2011.

\bibitem{sun2023microscopic}
X.-X. Sun, L.~Guo, {\em et~al.}, ``Microscopic study of fusion reactions with a
  weakly bound nucleus: Effects of deformed halo,'' {\em Physical Review C},
  vol.~107, no.~1, p.~L011601, 2023.

\bibitem{zhao2020microscopic}
J.~Zhao, T.~Nik{\v{s}}i{\'c}, D.~Vretenar, {\em et~al.}, ``Microscopic model
  for the collective enhancement of nuclear level densities,'' {\em Physical
  Review C}, vol.~102, no.~5, p.~054606, 2020.

\bibitem{roy2011study}
P.~Roy, K.~Banerjee, S.~Kundu, T.~Rana, C.~Bhattacharya, M.~Gohil,
  G.~Mukherjee, J.~Meena, R.~Pandey, H.~Pai, {\em et~al.}, ``Study of
  light-particle evaporation spectra (n, p, $\alpha$) in 4 he+ 93 nb
  reaction.,'' in {\em Proceedings of the DAE Symp. on Nucl. Phys}, vol.~56,
  p.~540, 2011.

\bibitem{mengoni1994fermi}
A.~Mengoni and Y.~Nakajima, ``Fermi-gas model parametrization of nuclear level
  density,'' {\em Journal of Nuclear Science and Technology}, vol.~31, no.~2,
  pp.~151--162, 1994.

\bibitem{schmidt2012inconsistencies}
K.-H. Schmidt and B.~Jurado, ``Inconsistencies in the description of pairing
  effects in nuclear level densities,'' {\em Physical Review C}, vol.~86,
  no.~4, p.~044322, 2012.

\bibitem{zelevinsky2018constant}
V.~Zelevinsky, S.~Karampagia, and A.~Berlaga, ``Constant temperature model for
  nuclear level density,'' {\em Physics Letters B}, vol.~783, pp.~428--433,
  2018.

\bibitem{dilg1973level}
W.~Dilg, W.~Schantl, H.~Vonach, and M.~Uhl, ``Level density parameters for the
  back-shifted fermi gas model in the mass range 40< a< 250,'' {\em Nuclear
  Physics A}, vol.~217, no.~2, pp.~269--298, 1973.

\bibitem{ignatyuk1993density}
A.~Ignatyuk, J.~Weil, S.~Raman, and S.~Kahane, ``Density of discrete levels in
  sn 116,'' {\em Physical Review C}, vol.~47, no.~4, p.~1504, 1993.

\bibitem{minato2011nuclear}
F.~Minato, ``Nuclear level densities with microscopic statistical method using
  a consistent residual interaction,'' {\em Journal of nuclear science and
  technology}, vol.~48, no.~7, pp.~984--992, 2011.

\bibitem{goriely2009first}
S.~Goriely, S.~Hilaire, M.~Girod, and S.~P{\'e}ru, ``First
  gogny-hartree-fock-bogoliubov nuclear mass model,'' {\em Physical review
  letters}, vol.~102, no.~24, p.~242501, 2009.

\bibitem{hilaire2012temperature}
S.~Hilaire, M.~Girod, S.~Goriely, and A.~J. Koning, ``Temperature-dependent
  combinatorial level densities with the d1m gogny force,'' {\em Physical
  Review C}, vol.~86, no.~6, p.~064317, 2012.

\bibitem{behkami2002level}
A.~Behkami, Z.~Kargar, and N.~Nasrabadi, ``Level density parameter study using
  a microscopic model,'' {\em Physical Review C}, vol.~66, no.~6, p.~064307,
  2002.

\bibitem{goodman1981finite}
A.~L. Goodman, ``Finite-temperature hfb theory,'' {\em Nuclear Physics A},
  vol.~352, no.~1, pp.~30--44, 1981.

\bibitem{egido1986temperature}
J.~Egido, P.~Ring, and H.~Mang, ``Temperature-dependent hartree-fock-bogoliubov
  calculations in hot rotating nuclei,'' {\em Nuclear Physics A}, vol.~451,
  no.~1, pp.~77--90, 1986.

\bibitem{goriely2008improved}
S.~Goriely, S.~Hilaire, and A.~J. Koning, ``Improved microscopic nuclear level
  densities within the hartree-fock-bogoliubov plus combinatorial method,''
  {\em Physical Review C}, vol.~78, no.~6, p.~064307, 2008.

\bibitem{chen2022three}
M.~Chen, T.~Li, B.~Schuetrumpf, P.-G. Reinhard, and W.~Nazarewicz,
  ``Three-dimensional skyrme hartree-fock-bogoliubov solver in coordinate-space
  representation,'' {\em Computer Physics Communications}, vol.~276, p.~108344,
  2022.

\bibitem{capote2009ripl}
R.~Capote, M.~Herman, P.~Oblo{\v{z}}insk{\`y}, P.~Young, S.~Goriely, T.~Belgya,
  A.~Ignatyuk, A.~J. Koning, S.~Hilaire, V.~A. Plujko, {\em et~al.},
  ``Ripl--reference input parameter library for calculation of nuclear
  reactions and nuclear data evaluations,'' {\em Nuclear Data Sheets},
  vol.~110, no.~12, pp.~3107--3214, 2009.

\bibitem{hauser1952inelastic}
W.~Hauser and H.~Feshbach, ``The inelastic scattering of neutrons,'' {\em
  Physical review}, vol.~87, no.~2, p.~366, 1952.

\end{thebibliography}

\bibliographystyle{ieeetr.bst}

\end{document}